# The Relationship between Foreign Direct Investment and Economic Growth: A Case of Turkey

Orhan Gokmen[1]

[1] M.A. in Applied Economics, Department of Economics, The George Washington University, USA

Correspondence: Orhan Gokmen, M.A. in Applied Economics, Department of Economics, The George Washington University, USA. ORCID ID: 0000-0002-9956-1179. E-mail: ogokmen@gwu.edu

## Abstract

This paper examines the relationship between net FDI inflows and real GDP for Turkey from 1970 to 2019. Although conventional economic growth theories and most empirical research suggest that there is  a bi-directional positive effect between these macro variables, the results indicate that there is a uni- directional significant short-run positive effect of real GDP on net FDI inflows to Turkey by employing the Vector Error Correction Model, Granger Causality, Impulse Response Functions and Variance Decomposition. Also, there is no long-run effect of net FDI inflows found on real GDP, yet vice-versa long-run effect has been found. The findings recommend Turkish authorities optimally benefit from the potential positive effect of net incoming FDI on the real GDP by allocating it for the productive sectoral establishments while effectively maintaining the country's real economic growth to attract further FDI inflows.

## 1. Introduction

Financial globalization has been one of the key factors contributing to the mitigation of rigid economic boundaries between countries by enhancing cross-country capital mobility for almost two centuries. However, the enhancement process of international capital mobility was not a smooth one due to interruptive political and economic episodes throughout the 19[th] and early 20[th] century. The collapse of the Bretton Woods System in 1971 and the Oil Shock of 1973-1974 greatly revitalized and increased the volume of international capital flows among countries, specifically from developed to developing countries, thanks to the innovations in communication technologies, promising domestic factors of the developing countries and macro-finance factors such as growth in Eurodollar market, bank lending and low-interest earnings at the financial centers (Kaminsky, 2004). Starting from the late 1990s, the capital flows  have gained a remarkable momentum till the present day (2020) but with much more volatility and boom-bust cycles.

Foreign Direct Investment (FDI) as a long-term capital flow, which consists of foreign portfolio or equity investment that make up more than 10% of the voting stock of an enterprise, has been generally considered as the most desirable type of capital flows by many policymakers due to its less speculative and development-focused features compared to its short-term counterparts. Therefore, most governments of the developing countries have undertaken institutional reforms and designed economic policies to give incentive to investors to attract inflows of FDI by advertising actual or potential economic growth of their country as a pull factor for FDI inflows. However, it has also been a topic of controversies and target of pundits in the field of economics and politics concerning its role in driving economic and societal growth. From an optimistic perspective, some assert that FDI encourages economic growth and productivity increase by transferring capital, knowledge, and technology to the host country with a low capital/labor ratio, i.e., the high marginal productivity of capital, which justifies the motivation of foreign investors and domestic business environment. Also, they underline the positive role of FDI in speeding up the "catch-up" process  with the developed country for the developing host country. On the other hand, others emphasize the risk of FDI damaging macroeconomic stability by overly appreciating domestic currency in the short-run, impeding domestic economic activity by harming local capabilities such as making domestic firms less efficient compared to foreign ones due to "market stealing effect", making host country more dependent on the home country, extracting natural resources without compensating the host country and, lastly, creating employment bias towards high-skilled workers due to R&D spillovers from FDI, which can contribute to unemployment in the low-skilled labor intensive country.

As a labor-intensive developing country, Turkey has experienced mostly similar paths with the rest of the World in the evolution timeline of capital inflows but with different magnitudes when compared to prominent Latin American and Caribbean developing countries. As shown in Figure 1,from 1970 to 1975, Turkey realized almost the same FDI net inflows (inflows minus outflows) as the World average. From 1976 to 1981, due to the political turbulence and military coup, FDI reached nearly 0% share of GDP level. Starting from 1987, thanks to both external and internal factors, Turkey enjoyed increased FDI net inflows. However, this increase in net flows was not sustained, and huge divergence between Turkey and the World average/Latin America became apparent in the 1990s due to negative domestic economic and political forces. After the 2000s, the volatility of FDI net inflows greatly increased due to imbalances in the global financial system and the ever-changing risk appetite of investors.

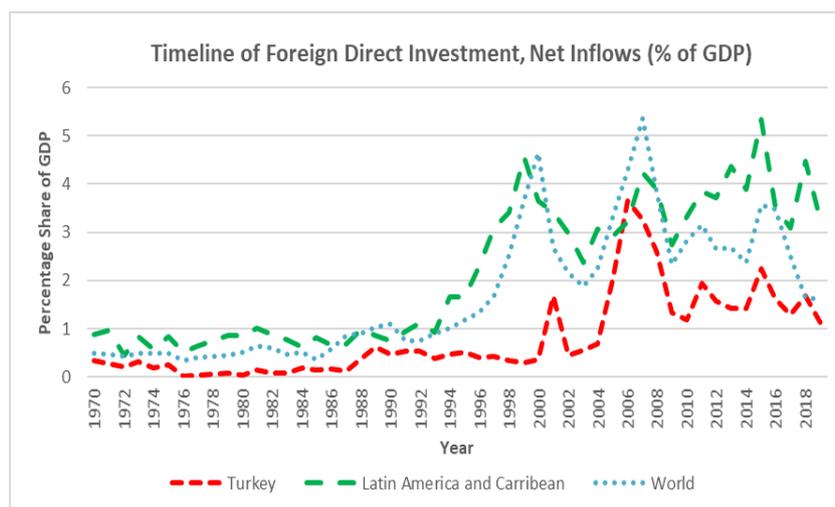

Figure 1. Evolution of FDI among Turkey, Latin America/Caribbean, and the World

Source: World Bank, World Development Indicators.

As a result of the paradox regarding the role of FDI in the economic growth of developing countries and pull factor of domestic economic growth for FDI inflows, this paper will try to answer the question: "*How does foreign direct investment affect the economic growth of Turkey and how do they feed each other?* " by employing Vector Error Correction Model (VECM) to analyze the nexus between FDI and the economic growth of Turkey.

## 2. Literature Review

The relationship between FDI and economic growth has attracted substantial attention from both professionals and scholars since the capital markets of the most developing countries liberalized and capital mobility was greatly facilitated in the 1980s. Since the early 1990s, extensive empirical research has focused on this topic by using single or multi-country data with various models and methodologies thanks to advances in econometric analysis techniques and data availability.

Among one of the early empirical evidence, Balasubramanyam et al. (1996) investigated the role of FDI in the growth process of 46 developing economies that were categorized by their trade policy regimes from 1970 to 1985. Using cross-section data, they employed ordinary least squares and generalized instrumental variable estimator methods for each country category: import-substitution (IS), export- promotion (EP) and all countries in the sample. They found a positive impact of FDI on both IS and EP countries. Yet, they argued that FDI had more growth-enhancing effect for countries with EP policy than countries that followed IS one. As a limitation of the research, they asserted that their categorization process of the developing countries regarding their trade policy regimes was not perfectly aligned with the World Bank's broad classification terminology. In the following years, Bende-Nabende and Ford (1998) found evidence of FDI's positive economic growth effect by regressing the annual data from 1959 to 1995 with the 3SLS estimation method for Taiwan. Moreover, Borensztein et al. (1998) supported this evidence by using a cross-country regression framework for 69 developing countries from 1970 to 1989. They also suggested that higher productivity of FDI holds only when the host country has enoughabsorptive capability of the advanced technologies.

Contrary to previous evidence, Mencinger (2003) used a sample of 8 European Union countries that were in a transition period between 1994-2001 to examine the effect of FDI on economic growth by using the Granger causality test and OLS regression. The author asserted that the main channel of FDI's contribution to economic growth is the spillover effect. As an ultimate finding, the study found a negative uni-directional effect of FDI on the economic growth of the sample of countries. Moreover, Herzer (2010)extensively investigated the impact of FDI on economic growth and country-specific factors explaining the cross-country variances in the growth effect of FDI. The study used a sample of 44 developing countries from 1970 to 2005 by employing a heterogeneous panel cointegration technique and a general-to-specific model selection approach. Consistent with the findings of Mencinger (2003), the conclusion of the study emphasized that FDI has a negative effect on the growth of developing countries, on average and domestic factors such as human capital, openness, financial market development cannot explain the growth effects of FDI. Yet, the growth effect is positively related to less government intervention and business regulation and negatively related to natural resource dependence and FDI volatility.

For more recent studies, Chaudhry et al. (2013) conducted their research to find out the short-term and long-term effects of FDI on the economic growth of China. They used the World Bank's annual time series data from 1985 to 2009 by employing the ARDL co-integration approach and Error Correction Model.The study chose variables of net FDI inflow, gross fixed capital formation, general government final consumption expenditure to explain the sample variances in gross domestic product. The findings of the study indicated a positive short-run and long-run relationship between FDI and economic growth in China. Furthermore, Siddikee and Rahman (2020) used the VECM approach to show the relationship between net FDI inflows and GDP by using annual data from 1990 to 2018 for Bangladesh. As a result,they found an insignificant effect of FDI on economic growth for the short and long run.

For the evidence from Turkey, Bildirici et al. (2010) used quarterly time series data from 1992:Q1 to 2008:Q4 to analyze the relationship between FDI and economic growth derived from Industrial Production Index. They employed threshold the VAR models and separated periods into two regimes. The extreme regime indicated the period of considerable decreases in FDI inflows, whereas the typical regime maintained these inflows almost constantly. As a conclusion of their research, they found positive effects of FDI inflows on economic growth in both regimes. However, as a limitation, the definition of the economic growth as Industrial Production Index rather than GDP can lead to inaccurate results in finding the real impact of FDI on economic growth. In the same year, Ilgun et al. (2010) collected annual data from 1984 to 2004 to examine the short-run effect of net FDI inflow by establishing VAR model. They utilized the Cobb-Douglas production function for their model and chose growth rate of GDP, net inflows of FDI, total labor force, gross fixed investment, and balance of payments as variables. As a conclusion, they found a positive bi-directional relationship between FDI and GDP. Also, their research showed that the main source of variance to GDP among variables was FDI followed by total labor force and condition of the balance of payment. Nevertheless, the authors argued that the research had a limitation stemming from a small number of observations. Finally, Demirsel et al. (2014) conducted their research considering quarterly time periods between 2002:Q1 and 2014:Q1 by using two variables: FDI inflows and GDP. They found no long-run relationship between those variables by only applying the Johansen cointegration test.

Taking mentioned research into consideration, the empirical evidence has not provided a consensus on this topic because the economic, political, and demographic structures of countries have created unique grounds with different magnitudes for FDI's impact on economic growth. In addition,the authors' different approaches in the selection of variable, timeline and model have caused various interpretations in the examination of the topic.

## 3. Theoretical Background

There have been two main underlying theories for the effect of FDI on economic growth. The first is the exogenous-growth theory, i.e., the neo-classical model that was developed by Solow and Swan (Solow, 1956 and 1957). The model, which was influenced by Cobb-Douglas production function (Cobb & Douglas, 1928), assumes that growth can be achieved by exogenous factors such as stock of capital, labor, and technology.

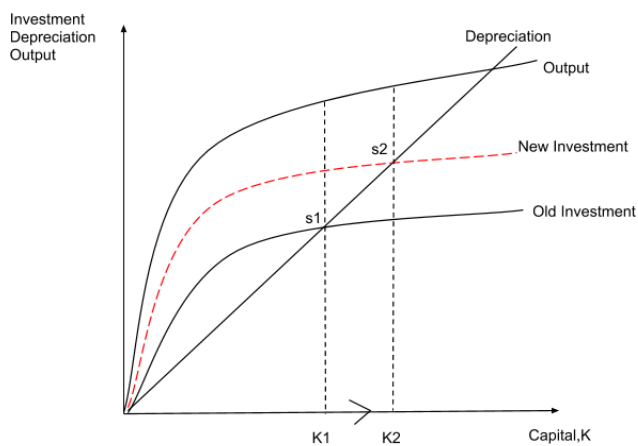

Figure 2. Neo-classical growth model

As shown in Figure 2, when there is an increase in the investment, that will increase both capital (from K1 to K2) and output, which also increases the economic growth and shifts the steady-state growthto a higher level (from s1 to s2). Moreover, positive externalities of FDI from the possible new technology adoption or transfer of know-how can further enhance the efficiency of investment in the hostcountry, positively affecting the economic growth. Ultimately, the neo-classical growth model indicates that FDI can affect economic growth of a country directly from capital accumulation, inputs, and technology channels.

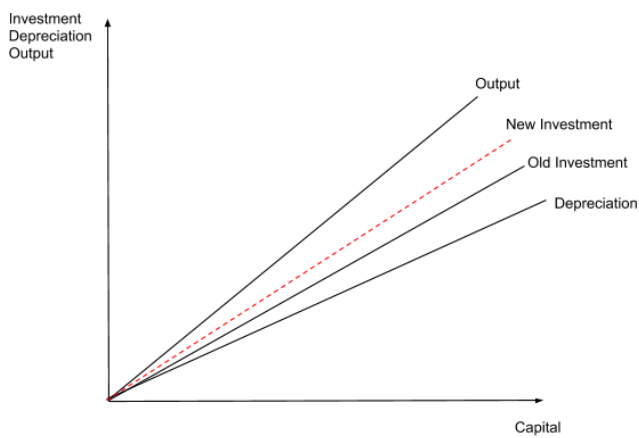

Figure 3. AK model

The second model concerning economic growth is the AK model, i.e., endogenous growth model, pioneered mainly by Romer and Lucas (Romer, 1986, 1990 and 1994; Lucas, 1988). Unlike the neo-classical growth model that considers technology as an exogenous factor, endogenous growth model assumes that economic growth is flourished by stock of human capital and technological changes, which implies that the technology is developed endogenously due to an increase in knowledge and innovation. Thus, the role of FDI in this model is that it can increase the rate of economic growth infinitely in the host country, given that technology transfer and diffusion take place as a spill-over effect of FDI. As Figure 3 shows, an increase in the investment leads to an increased gap between investment and depreciation that finally results in higher growth without the presence of a steady state.

Conditionally supporting the growth theories, the OECD (2002) asserted that FDI could create sustainable economic development, given the conditions of assisting in the formation of human capital, generating technology spillovers, increasing the host country's interaction with the global economy, and thriving the competition among firms in the business environment.

## 4. Data and Methodology

There can be numerous variables that have an impact on the economic growth of a country. To choose the appropriate combination of variables along with the interest variable, this research considered factors of availability for each variable from a reliable public data source and the power of degrees of freedom of the model. In this regard, this paper identified 5 variables and gathered annual time-series data between 1970-2019 for Turkey from the World Development Indicators provided by the World Bank. The interest variables are as follows:

*Real Gross Domestic Product (RGDP)*: The variable reflects the economic growth of the country. It is measured in 2010 constant U.S. dollars. Also, the logarithm of the variable (*LRGDP*) is taken as it is in large positive dollar value and to avoid potential heteroskedasticity.

*Net Foreign Direct Investment Inflows (FDI)*: The variable indicates the net foreign capital inflows (grossinflows minus outflows) to domestic business with ownership of at least 10% stock. It is the sum of equity and portfolio flows and measured as % of GDP.

The control variables are as follows;

*Domestic Credit to Private Sector (PRVT)*: It refers to financial resources provided to the private sector by financial institutions through loans, purchases of non-equity securities and trade credits. Therefore, it captures the domestic financial development and measured as % of GDP.

*Trade Openness (TRADE)*: The variable is the sum of the exports and imports of goods and services. Itimplies the integration degree of a country with global trade and measured as % of GDP.

*Government Consumption Expenditure (GOVCON)*: It includes all government current expenditures for purchases of goods/services, compensation of employees and military personnel. It is measured as % of GDP.

Table 1 shows the descriptive statistics of the variables that give crucial insights into the research data. Some variables have been showing more extreme behavior with outliers and are valued more in terms of GDP, such as *PRVT* and *TRADE*, compared to *FDI* and *GOVCON*. Overall, the table gives clues about the structural dynamics of the Turkish economy.

Table 1. Descriptive statistics

| Variable | Mean | Median | Max | Min | S.D. | Obs. |
|----------|--------|--------|--------|--------|--------|------|
| LRGDP | $26.78 | $26.77 | $27.85 | $25.72 | $0.62 | 50 |
| FDI | 0.80% | 0.44% | 3.65% | 0.02% | 0.86% | 50 |
| PRVT | 27.66% | 18.48% | 70.85% | 13.59% | 17.95% | 50 |
| TRADE | 36.46% | 38.84% | 61.39% | 9.10% | 14.74% | 50 |
| GOVCON | 11.94% | 12.18% | 15.77% | 7.52% | 2.10% | 50 |

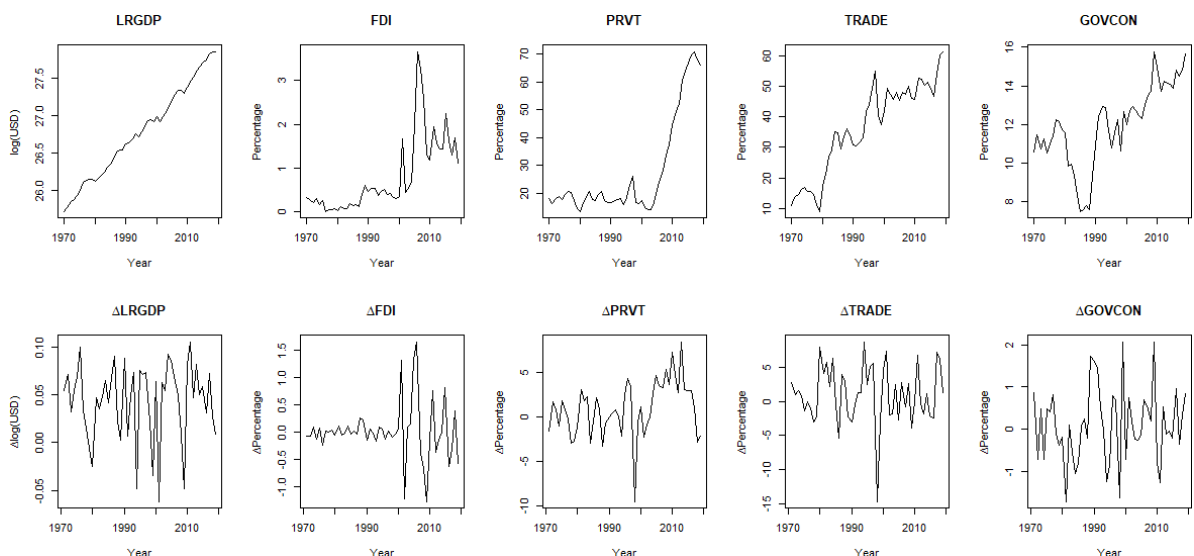

Figure 4. Time plot of variables at levels and first differences

Figure 4 indicates the time plots of each variable from 1970 to 2019. The effect of stochastic exogenous shocks is seen for each variable over time as they do not have constant mean, variance, and covariance. Almost all variables show an increasing trend, but to be more specific *FDI*, *PRVT* experienced a dramatic surge that occurred in the approximately same period (the early 2000s), which can be attributed to rise of confidence in the private sector and extensive financial and institutional development following the aftermath of the Turkish Financial Crisis in 2001 with the help of international monetary institutions. The second row shows that the trend effect is eliminated in their first differences. In addition, the mean, variance, and covariance of the variables become considerably stabilized compared to their values at levels. However, an increased higher variance is observed for *FDI* and *PRVT* starting from the early 2000s that reasserts the abnormal behavior in these variables at levels.

As seen previously in Figure 4, visually, the variables are non-stationary at their levels, but stationary at first differences and they are expected to influence each other in the context of the economic environment. Therefore, to investigate the dynamic relationship between FDI and economic growth of Turkey along with the other control variables, this paper employs VECM (Vector Error Correction Model), assuming there are cointegrating relationships between endogenous macro variables that show long-term stable relationship as proposed by Engle and Granger (1987), and further developed by Sims et al. (1990). Thus, there is a matrix system of the equation as shown in equation 1:

$$\begin{bmatrix} \Delta LRGDP \\ \Delta FDI \\ \Delta PRVT \\ \Delta TRADE \\ \Delta GOVCON \end{bmatrix}_t = \begin{bmatrix} \varphi_1 \\ \varphi_2 \\ \varphi_3 \\ \varphi_4 \\ \varphi_5 \end{bmatrix} + \sum_{k=1}^{n} \begin{bmatrix} \beta_{11} & \beta_{12} & \beta_{13} & \beta_{14} & \beta_{15} \\ \beta_{21} & \beta_{22} & \beta_{23} & \beta_{24} & \beta_{25} \\ \beta_{31} & \beta_{32} & \beta_{33} & \beta_{34} & \beta_{35} \\ \beta_{41} & \beta_{42} & \beta_{43} & \beta_{44} & \beta_{45} \\ \beta_{51} & \beta_{52} & \beta_{53} & \beta_{54} & \beta_{55} \end{bmatrix}_k \begin{bmatrix} \Delta LRGDP \\ \Delta FDI \\ \Delta PRVT \\ \Delta TRADE \\ \Delta GOVCON \end{bmatrix}_{t-k} + \begin{bmatrix} \gamma_1 \\ \gamma_2 \\ \gamma_3 \\ \gamma_4 \\ \gamma_5 \end{bmatrix} \begin{bmatrix} ECT_1 \\ ECT_2 \\ ECT_3 \\ ECT_4 \\ ECT_5 \end{bmatrix}_{t-1} + \begin{bmatrix} \varepsilon_1 \\ \varepsilon_2 \\ \varepsilon_3 \\ \varepsilon_4 \\ \varepsilon_5 \end{bmatrix}_t \quad (1)$$

Since the paper is interested in the relationship between FDI and GDP, the following equations 2 and 3 are the focus of the analysis:

$$\Delta LRGDP_t = \varphi_1 + \sum_{k=1}^{n} \beta_{11k} \Delta LRGDP_{t-k} + \sum_{k=1}^{n} \beta_{12k} \Delta FDI_{t-k} + \sum_{k=1}^{n} \beta_{13k} \Delta PRVT_{t-k} + \sum_{k=1}^{n} \beta_{14k} \Delta TRADE_{t-k}$$
$$+ \sum_{k=1}^{n} \beta_{15k} \Delta GOVCON_{t-k} + \gamma_1 ECT_{1(t-1)} + \varepsilon_{1t} \quad (2)$$

$$\Delta FDI_t = \varphi_2 + \sum_{k=1}^{n} \beta_{21k} \Delta LRGDP_{t-k} + \sum_{k=1}^{n} \beta_{22k} \Delta FDI_{t-k} + \sum_{k=1}^{n} \beta_{23k} \Delta PRVT_{t-k} + \sum_{k=1}^{n} \beta_{24k} \Delta TRADE_{t-k}$$
$$+ \sum_{k=1}^{n} \beta_{25k} \Delta GOVCON_{t-k} + \gamma_2 ECT_{2(t-1)} + \varepsilon_{2t} \quad (3)$$

where $\Delta$ is the first difference operator, $\varphi i$ are the constants, $\beta_{ijk}$ are the short-run coefficients, $\gamma i$ are the coefficients of the $ECT_{i,t-1}$, that are adjustment from short-term to long-term, $\varepsilon_i$ are disturbance terms that are assumed as a white-noise, t is the time, and k is the time lag length given for all i,j=1,2…5.

## 5. Empirical Results

The presence of the unit root in non-stationary time series variables can lead to spurious regression and lead to biased results. Therefore, this paper conducts Augmented Dickey-Fuller (ADF) and Phillips-Perron (PP) unit root tests to ensure the robustness of the model as all series need to be integrated of order 1 to employ the Vector Error Correction Model. Table 2 shows the results of the unit root tests.

Table 2. Results of unit root tests

| Variables | ADF | | | | PP | | | |
|---|---|---|---|---|---|---|---|---|
| | Level | | First difference | | Level | | First difference | |
| | T-stat | Crit. Val.(5pct) | T-stat | Crit. Val.(5pct) | T-stat | Crit. Val.(5pct) | T-stat | Crit. Val.(5pct) |
| LRGDP | -2.78 | -3.50 | 4.78* | -2.93 | -2.90 | -3.50 | -6.67* | -2.92 |
| FDI | -3.57* | -3.50 | 5.65* | -2.93 | -3.16 | -3.50 | -6.59* | -2.92 |
| PRVT | -1.30 | -3.50 | 3.40* | -2.93 | -0.89 | -3.50 | -4.42* | -2.92 |
| TRADE | -3.39 | -3.50 | 5.83* | -2.93 | -3.04 | -3.50 | -6.28* | -2.92 |
| GOVCON | -2.33 | -3.50 | 4.17* | -2.93 | -2.31 | -3.50 | -6.95* | -2.92 |

*Note.* * denotes rejection of the null hypothesis at 5% level of significance. Both unit root tests are conducted with considering combinations of trend and drift components appropriately, and max lag length is selected as 3 due to having annual data for both unit root tests.

According to the ADF test, all variables are integrated of order 1, that is I(1), as the null hypothesis of non-stationarity is rejected at first differences not at levels, except FDI, which is I(0) but the value of t-stat and critical value at 5 percent do not considerably differ from each other. For the PP test, all variables are I(1).

Although there is a contradiction between ADF and PP in the integration order of FDI, this paper takes the PP unit root test as a reference and concludes that all variables are I(1), which is also consistent with the previous visual inspection of the series' time plots and ACF correlograms of FDI (see Appendix A6- A7). Ultimately, the result justifies the employment of the Vector Autoregressive Models as all variables are integrated at the same order.

Table 3. VAR lag order selection criteria

| Lag | AIC | HQ | SC | FPE |
|---|---|---|---|---|
| 1 | -3.3680 | -2.9236 | -2.1870* | 0.0347 |
| 2 | -3.8867* | -3.0720* | -1.7217 | 0.0214* |
| 3 | -3.6642 | -2.4791 | -0.5150 | 0.0295 |

*Note*. * indicates lag order selected by the criterion with the minimum value.

Table 3 shows the optimal lag for the variables is 2, considering the minimum values of AIC, HQand FPE criteria for each lag. Meanwhile, the SC criterion indicates 1 lag. Therefore, as most of the criteria suggest 2 lags, it will be used for Johansen Cointegration Test and for the model to make it the most parsimonious one.

Table 4. Results of Johansen cointegration test

| Null Hypothesis | Eigen T-Statistic | Critical Val. (5pct) | Trace T-Statistic | Critical Val. (5pct) |
|---|---|---|---|---|
| r = 0 | 45.13* | 34.40 | 109.97* | 76.07 |
| r ≤ 1 | 34.50* | 28.14 | 64.84* | 53.12 |
| r ≤ 2 | 18.16 | 22.00 | 30.34 | 34.91 |
| r ≤ 3 | 8.00 | 15.67 | 12.18 | 19.96 |

*Note*. * denotes rejection of the null hypothesis at 5% significance level. The test is conducted with 2 lags and a constant term.

As shown in Table 4, the Johansen Cointegration test, which is to show if there is a common stochastic trend, stationarity that results from an identical linear combination or long-run relationship between I(1) variables, reveals that there are at most 2 cointegrating vectors since the null hypothesis of both Eigen and Trace tests, that is "no cointegration" and "at most 1 cointegration" are rejected at 5% significance level and the null of "at most 2 cointegration" is accepted. Thus, there is cointegration, that is, the presence of a long-run relationship between the variables besides the short-run. Eventually, cointegration with I(1) variables implies the error correction model. Employing VECM instead of simple VAR will provide a better understanding of the relationship dynamics among the variables and prevent the model estimation from being biased. More importantly, it sheds light on the long-run adjustment mechanism that shows the correction speed of the variables from their long-run disequilibrium by error correction term.

Table 5. Unrestricted VECM short-run estimation results with adjustment coefficients

| Variable Equation | ECT | ΔLRGDP(-1) | ΔFDI(-1) | ΔPRVT(-1) | ΔTRADE(-1) | ΔGOVCON(-1) |
|---|---|---|---|---|---|---|
| ΔLRGDP | -0.1083* | -0.2344 | 0.0234 . | 0.0065** | -0.0030 . | 0.0063 |
| ΔFDI | -0.4943*** | 3.5673 . | 0.2108 | 0.0421 | 0.0109 | -0.0050 |

*Note*. \*\*\*,\*\*,\*, . denotes 0%,1%, 5% and 10% percent significant levels,respectively.Estimated by Maximum Likelihood method.AIC= -193.26.

Table 5 shows the short-run coefficients with alphas (ECT) for the given variable equations.

Since the Johansen cointegration test results suggested two cointegrating vectors, this output uses the appropriate ECTs for the variables of interest. The equation *LRGDP* shows the error correction term as negative and significant at 5%, which means the previous deviation of GDP from long-run equilibrium is corrected by a speed of 11% each year, so in approximately 9 years, it will align with long-run equilibrium. It is seen that the real GDP has an insignificant negative effect on its growth for the following year. On the other hand, net FDI inflows and credit to private sector have a significant positive effect on next year's real GDP. Trade openness has a significant small negative effect, and government expenditure has an economically insignificant positive effect on real GDP for the next year. For the equation *FDI*, the error correction term is significant and negative as well, which means FDI is being corrected towards the long-run equilibrium due to influences from the right-hand side variables, but with an extraordinarily slow pace of adjustment. In the equation, only real GDP has a significant positive effect on next year's FDI. The impacts of current FDI, credit to private sector, trade openness on next year's FDI are insignificant positive, yet the effect of government consumption indicates insignificant negative effect. Moreover, it is useful to be suspicious about model misspecification that can stem from weakly exogenous variables, which are important for the short run but do not determine the long-run relationship among variables, because in a 5- variable model, the chance of having a weakly exogenous variable is very high and very slow convergence rates as shown by both error correction terms can be a sign of the misspecification. By testing the weak exogeneity for each variable by employing likelihood ratio test to avoid estimation biases in variable coefficients and to find the most parsimonious model (see Appendix A2), TRADE and GOVCON are found as weakly exogenous because the null of the restriction cannot be rejected at 10% significance level so they can be left out of cointegrating space and long-run relationship since the VECM framework assumes all variables as endogenous that are affected endogenously by each other in the model. Ultimately, Table 6 shows the restricted vector error correction model.

Table 6. Restricted VECM short-run estimation results with adjustment coefficients

| Variable Equation | ECT | ΔLRGDP(-1) | ΔFDI(-1) | ΔPRVT(-1) |
|---|---|---|---|---|
| ΔLRGDP | 0.0022 | -0.2261 | 0.0160 | 0.0067** |
| ΔFDI | -0.4786*** | 3.8425* | 0.2108 | 0.0421 |

*Note.* ***,**,* denotes 0.1%, 1% and 5% percent significant levels, respectively. Estimated by Maximum Likelihood method. AIC= -294.53.

Having the same model specifications as suggested by optimal lag selection and Johansen cointegration test (see Appendix A4-A5) and using proper ECTs, the restricted model seems to be more parsimonious with lower AIC value and significant negative error correction term for *FDI* equation implying, again, a very slow convergence pace, nearly 1% per year, to the long-run equilibrium from past short-run deviations of *FDI*. Interestingly, it is noticeable that the error correction term has become positive and insignificant for the *LRGDP* equation, which implies the presence of only a short-run relationship as there is equilibrium between short and long-run values, so the changes in real GDP completely adjust to changes in net FDI inflows and domestic credit to private sector in the short run as there is no disturbance effect present by the influence of the right-hand side variables. Holding ceteris paribus assumption for all interpretations; 1% increase in current real GDP will potentially decrease the next year's real economic growth by 0.23% without any statistical significance, this negative sign can imply the countercyclical monetary and fiscal policies that are implemented by Turkish policymakers to slow down the economic activity and to increase savings. Also, 1% increase in net FDI in flows will increase next year's real GDP by 1.6%, which is statistically insignificant, but a positive sign shows the potential stimulative effect of FDI even it does not have an actual impact on the Turkish economy. Furthermore, credit to private sector has both statistically and economically significant effect on the next year's real economic growth that emphasizes the driving role of financial development on Turkey's real GDP. For the equation *FDI*, a 1% increase in current real GDP will increase next year's net FDI inflows as a share of GDP by 0.04% with significance at 5% level that indicates the motivating role of the economic growth for foreign investors as an incentive to engage in direct investment. Moreover, the effect of net FDI inflows on its own value in the next year is positive but statistically insignificant that infers the potential herding effect among foreign investors in their investment decisions. Lastly, there is an insignificant positive effect of domestic credit to private sector on next year's net FDI inflows.

Table 7. Restricted VECM long-run estimation results

| Cointegrating Vector | FDI | LRGDP | PRVT |
|---|---|---|---|
| r1 | 0 | 1 | -0.0292 |
| r2 | 1 | 0 | -0.1748 |

Table 7 shows two vectors as recommended by the Johansen cointegration test. Although a 1% increase in credit to private sector is expected to increase real GDP by almost 3%, the long-run interpretation of the first cointegrating vector can be problematic since the insignificant ECT from Table 6 indicates that only a short-run relationship is present for the variables in the *LRGDP* equation. Whereas the second cointegrating vector shows that a 1% increase in domestic credit to private sector will increase net FDI inflows by 0.17% in the long run with statistical significance as the error correction term for *FDI* equation 2 is significant at all levels. Therefore, the normalized form of vector equation can be constructed by reversing the signs to interpret the long-run coefficients properly, as equation 4 shows;

$$FDI = 0.1748PRVT \qquad (4)$$

Also, it is important to consider that real GDP is not in the second cointegrating equation as it has a value of zero, so long-run behavior will be more precise while analyzing the IRFs and variance decomposition. Next, it is useful to know the direction of the causality between FDI and GDP by eliminating the effect of credit to private sector employing the Granger causality test. However, it is important to note that Granger causality does not reflect the actual effect of a variable on another variable with its limitations. In fact, this type of causality emphasizes the explanatory power of past values of a variable in helping another variable's predictability.

Table 8. Granger causality test

| Null Hypothesis | Lag | p-value | Decision |
|---|---|---|---|
| FDI does not Granger cause LRGDP | 2 | 0.7955 | Accept |
| LRGDP does not Granger cause FDI | 2 | 0.0189 | Reject |

As Table 8 shows, net FDI inflows does not improve the predictability of the real GDP of Turkey, which also means that it does not reduce the residuals of the model. Meanwhile, real GDP improves the predictability of the net FDI inflows to Turkey as the null hypothesis is rejected. Thus, the results suggest unidirectional Granger causality from real GDP to net FDI inflows consistent with the restricted VECM results.

Since the model results are derived, it is vital to check the robustness of the estimations by diagnostic tests before proceeding to impulse response functions and variance decomposition.

Table 9. Diagnostic tests

| ~ Values correspond to p-value~ | Unrestricted VECM (r=2) | Restricted VECM (r=2) |
|---|---|---|
| Portmanteau Autocorrelation Test | 0.07472 | 0.2804 |
| Jarque-Bera Normality Test | 0.03796 | 0.0008 |
| ARCH Test for Heteroskedasticity | 1 | 0.1076 |

Table 9 indicates the diagnostics of the model. There is no autocorrelation in the model's residuals as the Portmanteau test suggests the null of absence of serial correlation cannot be rejected at 5% level for both models. In addition, the ARCH test shows that the models do not suffer from ARCH effect, in other words, the model residuals do not exhibit conditional heteroskedasticity. Lastly, the residuals of the model are not normally distributed for both models, which can be due to outliers in the data such as a sharp increase in domestic credit to private sector in the early 2000s, excess skewness or kurtosis characteristics of the variables. After checking the robustness of the models, IRFs of the restricted VECM will better illuminate the periodic relationship between net FDI inflows and real GDP.

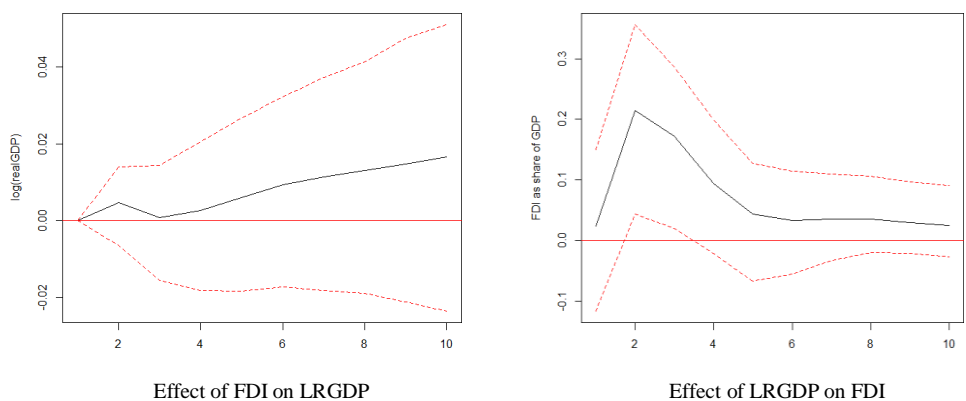

Effect of FDI on LRGDP          Effect of LRGDP on FDI

Figure 5. Impulse response functions for log. transformed Real GDP and Net FDI inflows (%GDP)

As shown in Figure 5, from the graph on the left, 1 standard deviation positive shock to net FDI inflows creates an immediate modest positive effect, which is insignificant as the effect is stuck in 1 interval, on real GDP till the second year, from the second year to the third, the gain in the real GDP got offset by a 1-year decline. Starting from third year, we see a continuous increase in the real GDP, constituting a similarity with the endogenous growth theory. For the graph on the right, a positive innovation shock to real GDP increases net FDI inflows quickly and significantly. The increase in net FDI inflows lasts for almost 1 year and hit the peak, then after the second year, the effect of real GDP on net FDI inflows starts to die out and from the fifth year, the effect becomes stabilized. The results show the economically feasible behavior of real GDP and net FDI inflows.

To further analyze the relationship between net FDI inflows and real GDP, variance decomposition tables (see Appendix A8 for visual version) can provide a good understanding of the interactions between the variables by showing the influence of a variable on the variation of another variable through the years. As Table 10 shows, the contribution of real GDP to the variance of itself declines gradually over the years. Net FDI inflows' contribution to variation in real GDP stays very minimal through time as a shock in FDI leads to 1% variation in real GDP in the second period, then its effect fades out till period 5. Starting from period 5, the contribution of FDI to variation of real GDP increases at a very slow pace, reaching to maximum effect by explaining a 4% variation in real GDP in period 9. Moreover, the effect of credit to private sector in explaining the variation in real GDP has much more effect than net FDI inflows with an increasing trend over the periods.

Table 10. Results of the Variance Decomposition for net FDI inflows and real GDP

| Period | Decomposition of LRGDP | | | Decomposition of FDI | | |
|---|---|---|---|---|---|---|
| | LRGDP | FDI | PRVT | FDI | LRGDP | PRVT |
| 1 | 1.00 | 0.00 | 0.00 | 1.00 | 0.00 | 0.00 |
| 2 | 0.91 | 0.01 | 0.09 | 0.84 | 0.15 | 0.01 |
| 3 | 0.86 | 0.00 | 0.13 | 0.75 | 0.19 | 0.05 |
| 4 | 0.84 | 0.00 | 0.16 | 0.72 | 0.20 | 0.07 |
| 5 | 0.82 | 0.01 | 0.17 | 0.72 | 0.20 | 0.08 |
| 6 | 0.80 | 0.02 | 0.19 | 0.72 | 0.20 | 0.08 |
| 7 | 0.77 | 0.02 | 0.20 | 0.71 | 0.21 | 0.08 |
| 8 | 0.75 | 0.03 | 0.21 | 0.71 | 0.21 | 0.08 |
| 9 | 0.73 | 0.04 | 0.23 | 0.71 | 0.21 | 0.08 |

Also, the results revealed that the contribution of FDI to its own variation declines instantly after period 1 but becomes almost stagnant starting from period 4. Meanwhile, the contribution of real GDP to the variation of FDI prompts a sharp increase by 15% from period 1 to 2, later, from period 2 to 3, variance contribution of real GDP increases by 4%, reaching 19%. Nevertheless, after the third period, the increasing trend loses its momentum by showing a 1% increase every three year until period 9, ending with a 21% variance contribution of real GDP to net FDI inflows. Finally, credit to private sector as an indicator of financial development can only contribute approximately 10% to the variance of net FDI inflows through time. Ultimately, the findings from variance decomposition are consistent with the previous results of this paper as net FDI inflow does not have a significant explanation for real economic growth, but real economic growth has a considerable implication for net FDI inflows to move into the Turkish markets.

## 6. Policy Implications and Conclusion

The domestic economic growth effect of FDI is mainly channeled through domestic capital, human capital, and technology stocks of a country. Firstly, FDI inflows increase the country's capital stock by increasing the deposits of the various types of banks, creating more credits for the private sectoral business activity, which is further expected to amplify the domestic investment. Secondly, it can bring new training programs and technology that can increase the efficiency and effectiveness of the labor and business activity, which, in turn, increases economic growth. Because of the economic growth, FDI inflows are pulled to the host's domestic economy. Thus, creating a feedback loop between FDI inflows and the economic growth of a country.

However, the research results indicated a one-way positive effect from economic growth to net FDI inflows for Turkey. This supports the view that Turkey has not effectively transformed net FDI inflows as a significant driver of economic growth. In fact, the most net FDI inflows have been achieved through ownership changes, i.e., mergers and acquisitions of less productive sectoral enterprises such as shopping malls, leisure/entertainment, and financial firms without contributing to the emergence of innovations and production. In the light of this knowledge, Turkish policymakers should redirect the incoming FDI into productive sectors to enhance the real economic activity via consumption, investment, export/import channels while minimizing the FDI outflows to keep the net FDI inflows high by increasing the quality of political and economic institutions, providing various incentives, and keeping domestic real economic growth strong.

This study investigates the relationship between net FDI inflows and real economic growth in Turkey by employing the Vector Error Correction Model for the annual series from 1970 to 2019. Despite the conventional economic growth theories, the empirical findings show a positive but insignificant effect of net FDI inflows on real GDP and a small significant positive effect of real GDP on net FDI inflows in the short run. Moreover, no long-run effect of net FDI inflows and domestic credit to private sector found on real GDP, yet the vice-versa long-run effect is found. These results are further supported by uni-directional Granger causality from real GDP to net FDI inflows, IRFs and Variance Decomposition. This paper is consistent with the empirical results of Siddikee and Rahman (2020) for the insignificant effect of FDI on GDP and Demirsel et al. (2014) for the insignificant long-run effect from FDI to GDP. Importantly, it also supports the conditional growth effect of FDI as asserted by OECD (2002). As a limitation of this study, there is non-normality detected in the model residuals that can be attributed to the presence of outliers, skewness, and kurtosis characteristics of the series over time. Therefore, further research can overcome this limitation by using high-frequency data with more observation points when high-frequency data become available to conduct a fully robust analysis for the case of Turkey in the future.

## References

Balasubramanyam, V., Salisu, M., &Sapsford, D. (1996). Foreign Direct Investment and Growth in EPand IS Countries. *The Economic Journal, 106*(434), 92-105. https://doi.org/10.2307/2234933

Bende-Nabende, A., & Ford, J. L. (1998). FDI, policy adjustment and endogenous growth: Multiplier effects from a small dynamic model for Taiwan, 1959–1995. *World Development, 26*(7), 1315-1330. https://doi.org/10.1016/S0305-750X(98)00043-6.

Bildirici, M., Alp, E. A., & Kayıkçı, F. (2010). Effects of Foreign Direct Investment on Growth in Turkey. *International Conference on Eurasian Economies 2010, Session 3B: Growth & Development*. https://doi.org/10.36880/C01.00149

Borensztein, E., De Gregorio, J., & Lee, J. W. (1998). How does foreign direct investment affect economic growth? *Journal of International Economics, 45*(1), 115-135. https://doi.org/10.1016/S0022-1996(97)00033-0

Chaudhry, N. I., Mehmood, A., & Mehmood, M. (2013). Empirical relationship between foreign direct investment and economic growth: An ARDL co-integration approach for China. *China Finance Review International, 3*. https://doi.org/10.1108/20441391311290767.

Cobb, C. W., & Douglas, P. H. (1928). A theory of production. *The American Economic Review, XVIII*.

Demirsel, M. T., Öğüt, A., & Mucuk, M. (2014). The Effect of Foreign Direct Investment on EconomicGrowth: The Case of Turkey. *Proceedings of International Academic Conferences 0702081*, International Institute of Social and Economic Sciences.

Engle, & Granger. (1987). Co-integration and Error Correction: Representation, Estimation, and Testing. *Econometrica, 55*(2), 251-276. https://doi.org/10.2307/1913236

Herzer, D. (2010). How does foreign direct investment really affect developing countries' growth? *IAIDiscussion Papers, No. 207*, Georg-August-Universität Göttingen, Ibero-America Institute for Economic Research (IAI), Göttingen

Ilgun, E., Koch, K., & Orhan, M. (2010). How Do Foreign Direct Investment and Growth Interact in Turkey? *Eurasian Journal of Business and Economics, 3*, 41-55.

Kaminsky, G. (2004). International Capital Flows: Blessing or Curse?. *Revue d'Économie du Développement, 12*(3), 83-119. https://doi.org/10.3917/edd.183.0083.

Lucas, R. (1988). On the mechanics of economic development. *Journal of Monetary Economics, 22*, 3-42. https://doi.org/10.1016/0304-3932(88)90168-7

Mencinger, J. (2003). Does Foreign Direct Investment Always Enhance Economic Growth? *Kyklos, 56*, 491-508. https://doi.org/10.1046/j.0023-5962.2003.00235.x

OECD. (2002). *Foreign direct investment for development: Maximising benefits, minimising Costs*. OECD, Paris.

https://doi.org/10.1787/9789264199286-en

Romer, P. M. (1986). Increasing returns and long run growth. *Journal of Political Economy, 94*, 1002-1037. https://doi.org/10.1086/261420

Romer, P. M. (1990). Endogenous technological change. *Journal of Political Economy, 98*, S71-S102. https://doi.org/10.1086/261725

Romer, P. M. (1994). The origin of endogenous growth. *Journal of Economic Perspectives, 8*(1), 3-22. https://doi.org/10.1257/jep.8.1.3

Siddikee, N., & Rahman, M. M. (2020). Foreign Direct Investment, Financial Development, and Economic Growth Nexus in Bangladesh. *The American Economist*, 1-16. https://doi.org/10.1177/0569434520938673

Sims, C. A., Stock, J. H., & Watson, M. W. (1990). Inference in Linear Time Series Models with Some Unit Roots. *Econometrica, 58*(1), 113-144. https://doi.org/10.2307/2938337

Solow, R. (1956). A contribution to the theory of economic growth. *Quarterly Journal of Economics, 70*, 65-94. https://doi.org/10.2307/1884513

Solow, R. (1957). Technical change and the aggregate production function. *The Review of Economics and Statistics, 39*(3), 312-320. https://doi.org/10.2307/1926047

## Appendix A

Table A1. Unrestricted VECM short-run estimation results with adjustment coefficients

| Variable Equation | ECT1 | ECT2 | $\Delta$LRGDP(-1) | $\Delta$FDI(-1) | $\Delta$PRVT(-1) | $\Delta$TRADE(-1) | $\Delta$GOVCON(-1) |
|---|---|---|---|---|---|---|---|
| $\Delta$LRGDP | -0.1083* | -0.0165 | -0.2344 | 0.0234 . | 0.0065** | -0.0030 . | 0.0063 |
| $\Delta$FDI | 1.6588** | -0.4943*** | 3.5673 . | 0.2108 | 0.0421 | 0.0109 | -0.005 |
| $\Delta$PRVT | 0.2513 | 1.6601* | -22.1986 . | -1.1080 | 0.4468** | -0.1397 | -0.8234 . |
| $\Delta$TRADE | 18.6577*** | -1.7826 | -6.0500 | -0.7869 | 0.3415 | 0.3789* | -0.8510 |
| $\Delta$GOVCON | -0.3123 | 0.5038* | 5.5521 | 0.1209 | -0.0948 . | -0.0584 | -0.0030 |

Table A2. Test of Weak Exogeneity

| Variable | Likelihood Ratio T-stat | P-value |
|---|---|---|
| LRGDP | 17.71 | 0.00 |
| FDI | 20.56 | 0.00 |
| PRVT | 4.74 | 0.09 |
| TRADE | 3.59 | 0.17 |
| GOVCON | 4.40 | 0.11 |

Table A3. Restricted VECM Short-Run Estimation Results with Adjustment Coefficients

| Variable Equation | ECT1 | ECT2 | $\Delta$LRGDP(-1) | $\Delta$FDI(-1) | $\Delta$PRVT(-1) |
|---|---|---|---|---|---|
| $\Delta$LRGDP | 0.0022 | -0.0172 | -0.2261 | 0.016 | 0.0067** |
| $\Delta$FDI | 0.6404** | -0.4786*** | 3.8425* | 0.2108 | 0.0421 |
| $\Delta$PRVT | 0.0598 | 1.5714* | -12.6510 | -1.2772 | 0.3272* |

Table A4. VAR Lag Order Selection Criteria

| Lag | AIC | HQ | SC | FPE |
|---|---|---|---|---|
| 1 | -5.7377 | -5.5599 | -5.2653* | 0.0032 |
| 2 | -6.0618* | -5.7507* | -5.2352 | 0.0023* |
| 3 | -5.7288 | -5.2844 | -4.5478 | 0.0033 |

Table A5. Results of Johansen Cointegration Test

| Null Hypothesis | Eigen T-Statistic | Critical Val. (5pct) | Trace T-Statistic | Critical Val. (5pct) |
|---|---|---|---|---|
| r = 0 | 31.42* | 22.00 | 64.84* | 34.91 |
| r≤1 | 29.43* | 15.67 | 33.42* | 19.96 |
| r≤2 | 3.98 | 9.24 | 3.98 | 9.24 |

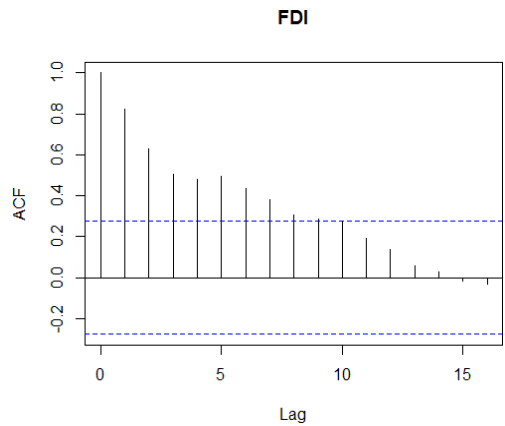

Figure A6. ACF correlogram of FDI at levels

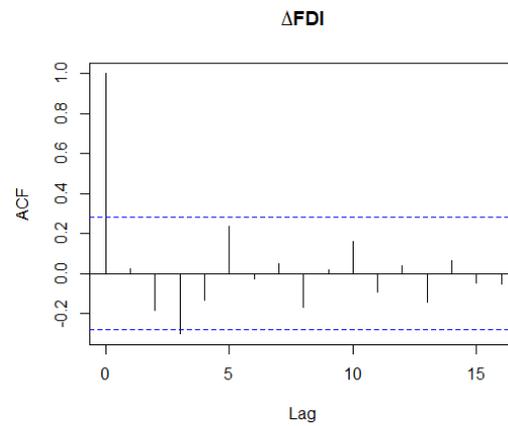

Figure A7. ACF correlogram of FDI at first differences

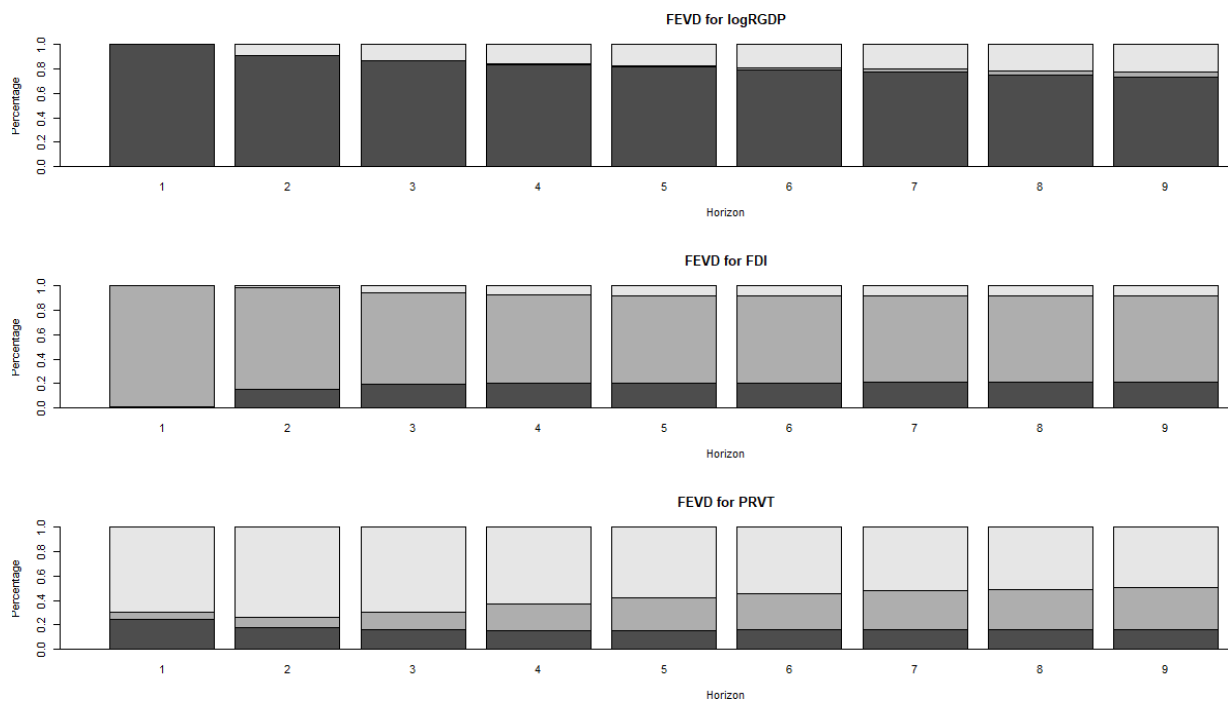

Figure A8. Visual Variance Decompositions

*Note.* Dark Grey represents logRGDP, Moderate Grey represents FDI, Light Grey represents PRVT.

## Copyrights